# Reduced Optimal Power Flow Using Graph Neural Network

Thuan Pham
*Student Member, IEEE*
Department of Electrical and Computer Engineering
University of Houston
Houston, TX, USA
tdpham7@cougarnet.uh.edu

Xingpeng Li
*Senior Member, IEEE*
Department of Electrical and Computer Engineering
University of Houston
Houston, TX, USA
xli82@uh.edu

*Abstract*— OPF problems are formulated and solved for power system operations, especially for determining generation dispatch points in real-time. For large and complex power system networks with large numbers of variables and constraints, finding the optimal solution for real-time OPF in a timely manner requires a massive amount of computing power. This paper presents a new method to reduce the number of constraints in the original OPF problem using a graph neural network (GNN). GNN is an innovative machine learning model that utilizes features from nodes, edges, and network topology to maximize its performance. In this paper, we proposed a GNN model to predict which lines would be heavily loaded or congested with given load profiles and generation capacities. Only these critical lines will be monitored in an OPF problem, creating a reduced OPF (ROPF) problem. Significant saving in computing time is expected from the proposed ROPF model. A comprehensive analysis of the GNN model's predictions was also made. It is concluded that the application of GNN for ROPF is able to reduce computing time while retaining solution quality.

## I. INTRODUCTION

Optimal power flow (OPF) is widely used for power system operations, analysis, and scheduling including real-time economic dispatch. OPF problems involve the optimization of an objective function that can take various forms while satisfying a set of operational and physical constraints. Applications of OPF have expanded because of its capabilities to deal with diverse aspects of the power system network. Due to the fast increase in renewable energy and its intermittent nature, total available generation capacity can fluctuate widely throughout the day. OPF calculation must be done rapidly and accurately to meet given demands while satisfying reliability of the system. Thus, improvements must be made upon the existing OPF model to meet the emerging challenging problems in power systems.

Different applications of machine learning (ML) in power system optimization have been studied extensively. For example, using recursive neural networks, researchers had tried to forecast energy prices of renewable energy sources for day-ahead energy market [1]; similarly, an integrated long-term recurrent convolutional network (LRCN) model was developed to forecast the wholesale electricity price [2]. The LRCN model is also utilized in [3] to estimate power system synchronous inertia. A multilayer neural network (NN) was proposed in [4] to predict power flow results as an alternative to traditional approaches. A logistic regression-based ML method is developed in [5] to determine a subset of generators' commitment status and thus reduce the day-ahead scheduling model. In [6], a NN-based battery degradation model is proposed for microgrid energy management.

Research in using ML for OPF has also been studied. In [7], an NN model was used to predict generations from load inputs in a security-constrained direct current OPF model. In [8], a convolutional neural network (CNN) model was built to predict generation dispatch using given load profiles. However, both methods do not account for topology of the power network. There is inherent information related to the global context of the network that can be used to train and enhance the ML model. Network topology is crucial and correlate to how well a learning model performs under different conditions [9].

In this paper, we proposed an ML algorithm using graph neural network (GNN) to assist efficient formulation of the OPF problem. GNN is an advanced NN model that captures the dependence of graphs via message passing between the nodes of graphs. GNN model relies heavily on network topology to pass information from nearby nodes and edges across multiple GNN layers. In [10], a comprehensive overview of GNN in power systems was reviewed in detail for power system applications such as fault detection, time-series prediction, and power flow calculation. Evaluations to use different variants of GNN for research has also been proposed [11]. Recently, GNN has been used to predict wind speed for generation at wind farm cluster [12]. GNN has the potential to expand and improve upon existing tools for power system optimization such as OPF.

A recent attempt to use GNN for OPF calculation did not account for power flow and ignore the line rating limit in its formulation of the OPF problem [13]. For this paper, we designed a GNN model that can predict overloaded and congested lines in the power network with different given load profiles. Predictions from the GNN model are used to identify a subset of critical lines for supervision. We want to lower the number of monitoring lines or number of line flow limit constraints in the OPF problem, in essence, producing a reduced

OPF (ROPF) problem. With an ROPF model, it is expected that the amount of computing time for finding the optimal solution will decrease, especially for large and complex power systems. Model selection and performance maximization of the proposed GNN model are performed in this paper. A GNN model was trained and tested against typical ML algorithms, such as NN and CNN, to assess and evaluate its overall accuracy.

The rest of the paper is organized as follows. Background analysis of optimal power flow and GNN are covered in section II. Section III introduces how sample data were generated, and the GNN model was developed. Section IV provides analysis of the results and evaluates the performance of the proposed GNN model. Section V concludes the paper and section VI describes possible future work.

## II. PRELIMINARIES

### A. Optimal Power Flow

OPF model is often applied for real-time economic dispatch in practical power systems. It is used to optimize generation dispatch with given system loads while satisfying multiple constraints throughout the transmission network. The objective function in OPF usually focuses on minimizing operating costs. The OPF model below also minimizes total system generation cost in (1) with variables of generation $P_g$, and the associated cost $c_g$ while it is subject to constraints (2)-(5).

$$\boldsymbol{Obj}: min \ \sum_{g\in G} c_g P_g \qquad g \in G \qquad (1)$$
$$P_g^{min} \leq P_g \leq P_g^{max} \qquad g \in G \qquad (2)$$
$$P_k = (\theta_{f(k)} - \theta_{t(k)})/x_k \qquad k \in K \qquad (3)$$
$$-RateA_k \leq P_k \leq RateA_k \qquad k \in K \qquad (4)$$
$$\sum_{g\in G(n)} P_g + \sum_{k\in K(n-)} P_k - \sum_{k\in K(n+)} P_k = d_n \quad n \in N \qquad (5)$$

Equation (2) sets the minimum/maximum generation limit of each generator in the system. The line flow on each line, $P_k$, is determined in (3) and is constrained by the line limit rating, $RateA_k$, in (4). $\theta_{f(k)}$ and $\theta_{t(k)}$ are the phase angles of the from-bus and to-bus of line $k$ respectively. Equation (5) denotes the nodal power balance with given nodal load, $d_n$.

For most OPF problems, the optimal solution is constrained by certain congested lines. These lines restrict the optimal solution within certain bounds and should be closely monitored. For most independent system operators, the constraint for line limit rating will only focus on a certain subset of lines that are heavily loaded or congested. These subsets of lines are chosen based on past solutions and past load profiles.

Our research focuses on how we can select these subsets of critical lines dynamically with respect to a given load profile using machine learning model, especially GNN. The GNN model should accurately predict which lines would be heavily loaded or congested given certain load profiles; then we only need to monitor those recognized critical lines rather than all lines in the OPF problem, which leads to an ROPF model relieving real-time computational burden. In summary, instead of using (4) to set line limit constraint for all lines, we will use (4a), where $r$ belongs to $R$, a subset of critical lines to monitor.

$$-RateA_r \leq P_r \leq RateA_r \qquad r \in R \qquad (4a)$$

Using predictions from GNN model, we establish an ROPF model to determine the best combination of generation outputs to meet the given load with minimal cost while maintaining line limit and nodal power balance constraints. The GNN model may be updated in near-real-time with different load profiles for different scenarios. It will allow system operators to achieve optimal generation dispatch within allocated computing time.

### B. Graph Neural Network

ML is an established computer algorithm that has been researched and applied toward power system optimization [5]. ML models can automatically improve/learn through experience using a large amount of data. Models are trained using sample training data set to make predictions or decisions without being explicitly programmed. CNN model and NN model are popular ML models that have been researched and used extensively for multiple applications across multiple disciplines. However, both models do not account for topology of the network in their predictions. For most electrical network, topology contains crucial, inexplicit information/features that can be extracted/learned to enhance the learning model. GNN model has an advantage compared to the other two models due to its usage of adjacency matrix (*nb* by *nb* matrix, where *nb* denotes number of buses) to describe topology of the network. In addition, the global context of the network itself is used to develop relationship between multiple features of nodes and edges. In short, a GNN model is an optimizable transformation on all attributes of the graph (nodes, edges, global context) that preserves graph symmetries (permutation invariances).

In Fig. 1 below, the bus-branch model of a simple power system was transformed into graph format. Each bus is a node with information such as generation and load represented as node features. Line flow and line thermal limit rating are edge features. The adjacency matrix characterizes the connection between each bus and its neighbors.

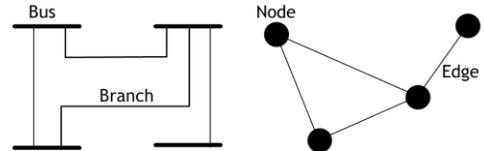

Fig. 1. Transformation of bus-branch model into graph format.

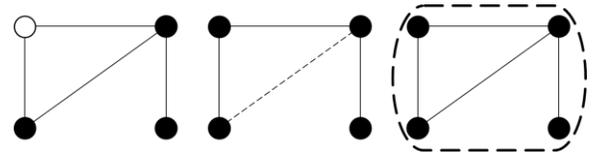

Fig. 2. Node-level vs. Edge-level vs. Graph-level classification task.

GNN has been used for multiple purposes. At the graph-level tasks, GNN is used to determine a single property for the entire graph. For example, GNN has been used extensively to predict the property of protein molecules based on their molecular structure, chemical bond, and polarity [14]. Similarly, for edge-level tasks, we want to identify the existence of an association between nodes. In citation networks, we explore the connections between research papers. We want to verify which of these nodes (research papers) share an edge (citation) or what the value of that edge is (number of citations).

## III. MODEL SELECTION

To generate a large number of samples to sufficiently train and test a GNN model, we run normal OPF simulations on the IEEE 73-bus system with different load profiles to collect 20,000 samples. The load profile of each sample is varied within ±10% of the base load profile. From the 20,000 generated samples, the data set is divided into three groups: 80% for training, 10% for validation, and 10% for testing. Then, we used Pyomo library in python to solve for the optimal solution of each sample [15] [16]. Based on the solution, we created labels for each branch based on a pre-specified loading threshold that is defined as a percentage of the line rating limit. For example, if the line rating limit is 100MW and the loading threshold is 80%, a branch will be classified/labeled as heavily loaded/congested when its flow is over 80MW. The labels are represented as one-hot encoding and used as output during the training process.

For each sample, the node features are the nodal load, maximum and minimum generation at each bus, the number of branches (edges) connected to each bus, and bus types (load bus, generator bus and slack bus). The edge features include the line reactance and the line rating limit.

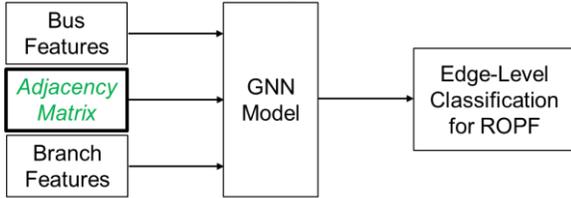

Fig. 3. Illustration of the GNN model. (4 XENet layers, 1 Dense layer)

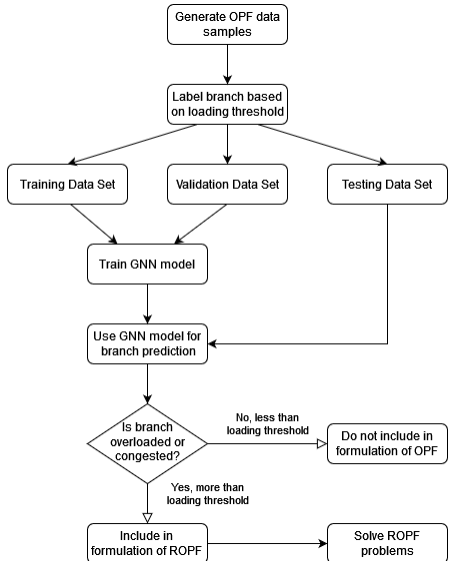

Fig. 4. Flow chart for constructing the proposed ROPF model using GNN.

As shown in Fig. 3, the proposed GNN model has four XENet layers followed by a dense layer. We decided to use XENet layer from the Spektral library as the primary layer to build the GNN model [17]. XENet layer is one of several popular GNN variants that has been developed recently for edge-level tasks. Using XENet, features from both nodes and edges are stacked and passed through each layer concurrently [18]. In addition, the adjacency matrix is passed along through each layer to retain the network topology and global context of the network during training. The final output is passed through a dense layer for edge-level classification. Once classification has been made, branches that are over the loading threshold will be include in the proposed ROPF model. The flowchart in Fig. 4 summarizes how ROPF is formulated using information predicted by the GNN model.

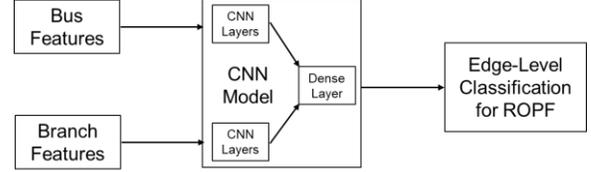

Fig. 5. Illustration of the CNN model. (4 CNN layers for bus features, 4 CNN layers for branch features, and 1 Dense layer to combine both features)

To evaluate how the GNN model perform against typical NN and CNN models, we built separate NN and CNN models that mimic the GNN model. For these models, node features and branch features are each passed and trained separately through four layers. Then, the outputs are combined in a dense layer with softmax activation. As we can see from Fig. 5 and Fig. 6, there is no adjacency matrix to keep track of network topology or global context during the training process.

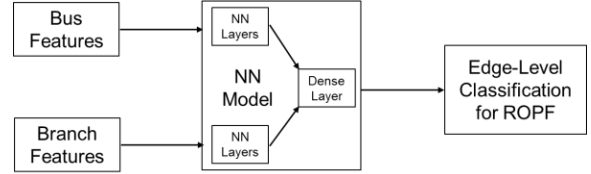

Fig. 6. Illustration of the NN model. (4 NN layers for bus features, 4 NN layers for branch features, and 1 Dense layer to combine both features)

## IV. CASE STUDIES

In this section, we evaluated how well the GNN model performed when we set the threshold at 95% of line rating limit for classifying whether a line is heavily loaded or congested. Model training and OPF computation were done on a system with 4 cores i7-4790K CPU, GTX970 graphic card, and 32GB memory.

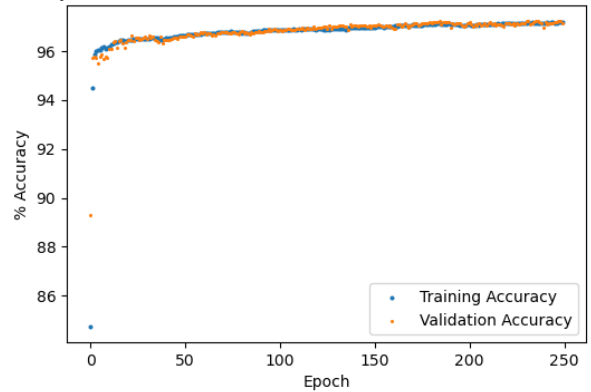

Fig. 7. Percent accuracy of prediction for training data set vs. validation data set during the training phase of GNN model.

From Fig. 7 above, we can see that the GNN model performs well during the training stage of 250 epochs. The training accuracy level is tracked consistently with the validation accuracy level.

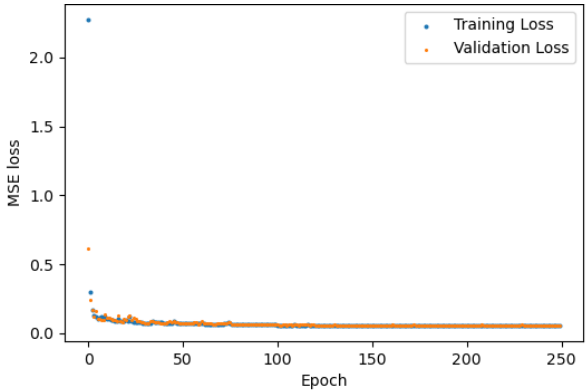

Fig. 8. MSE loss for training data set vs. validation data set during the training phase of GNN model.

As the training progress, the MSE loss dropped significantly within the first few epochs, then flattened and eventually saturated for the remaining epochs in Fig. 8.

With a trained GNN model, it was used to make predictions against the testing data set of 2,000 samples. From Fig. 9 to Fig. 13, we analyzed predictions of GNN model with a loading threshold for branch classification set at 95%. For most of the samples in Fig. 9, there are wrong predictions on at least one branch to as many as eight branches.

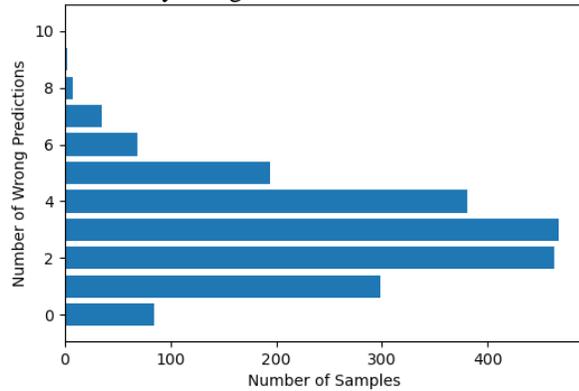

Fig. 9. Histogram represents the distribution of number of wrong predictions.

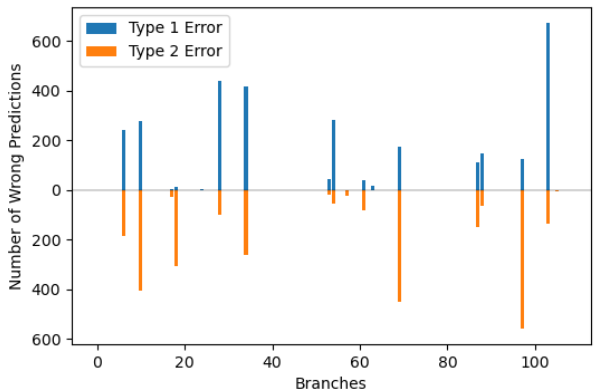

Fig. 10. Number of wrong predictions for Type 1 vs. Type 2 error per branch.

For each branch of each sample, prediction is always made on whether the branch is congested or not. Thus, we would have two possible errors: (i) "false positive/type 1" error – incorrectly predicting a lightly loaded line to be heavily loaded, and (ii) "false negative/type 2" error – incorrectly predicting a heavily loaded line to be lightly loaded. From Fig. 10 above, we separated incorrect predictions into two separate types. Then, we went further by visualizing the total number of wrong predictions for each branch based on the type of error. We noted that most of the prediction errors occur in a small number of branches.

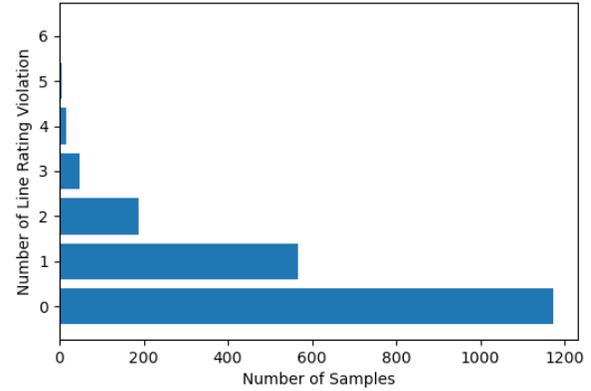

Fig. 11. Number of samples with at least 1 branch that violate line rating limit.

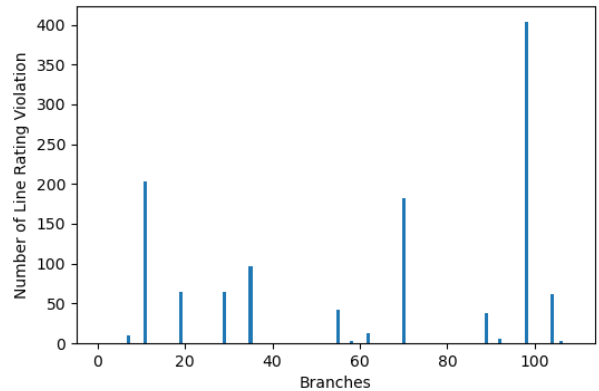

Fig. 12. Number of lines rating violation per branch.

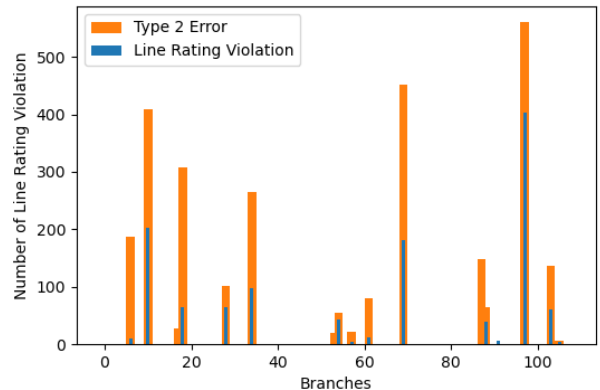

Fig. 13. Line rating violation is superimposed on top of type 2 error per branch.

Once predictions have been made, we ran ROPF calculations that only monitor branches that have been identified as congested by the GNN model. From the ROPF solutions in Fig. 11, it was noted that a significant number of samples contain unmonitored branches have violated the line rating limit. We then broke down the total number of line rating violations per branch in Fig. 12.

For type 2 error, we misclassified heavily loaded/congest lines. In turn, we did not include these lines in the group of monitored lines. Since these lines are critical, they are more likely to break the line limit constraint. In Fig. 13 above, Fig. 12 is superimposed on top of type 2 error in Fig. 10. We noticed a remarkable correlation, all branches that violate the line limit rating come from type 2 error almost exclusively

In Fig. 14, we shown the percent error prediction for different ML models at multiple thresholds. By setting multiple loading thresholds for branch classification, we wanted to determine which threshold would produce the lowest percent error in prediction. Additionally, we wanted to test how well the GNN model performed compared to both NN and CNN model. Based on the data on TABLE I, GNN model outperformed both NN and CNN model at every point of the loading threshold for branch classification. We also observed that there is a linear trend in which higher loading threshold will likely lead to higher prediction error overall across all ML models.

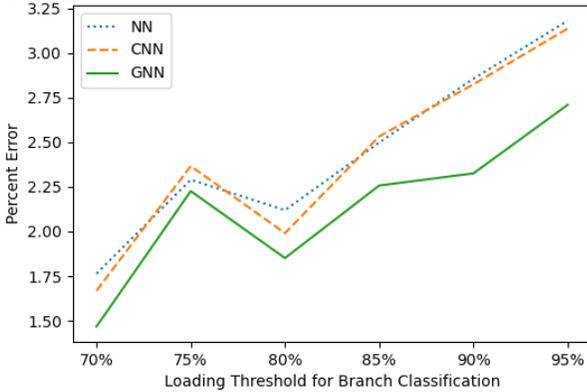

Fig. 14. Percent error of prediction at multiple loading thresholds from different ML models.

TABLE I
Percentage prediction error for different models across multiple loading thresholds for branch classification.

|     | 70%  | 75%  | 80%  | 85%  | 90%  | 95%  |
|-----|------|------|------|------|------|------|
| NN  | 1.76 | 2.29 | 2.12 | 2.50 | 2.86 | 3.18 |
| CNN | 1.67 | 2.37 | 1.99 | 2.53 | 2.82 | 3.14 |
| GNN | 1.47 | 2.23 | 1.85 | 2.26 | 2.33 | 2.71 |

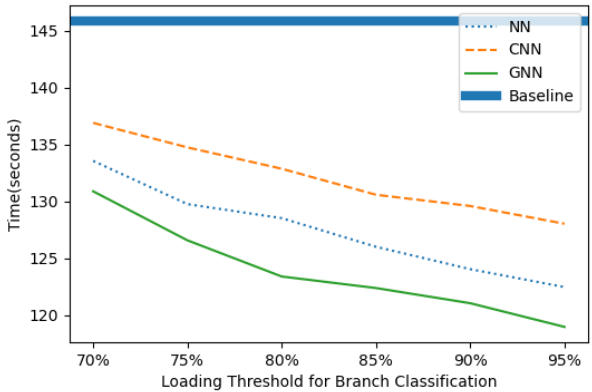

Fig. 15. Total time to solve 2000 ROPF testing samples at multiple loading threshold using predictions from different ML models.

From Fig. 14 and TABLE I, we can conclude that setting a high threshold for classifying congested lines will likely lead to a higher percentage of prediction error from the GNN model. With a high number of prediction errors, especially type 2 errors, there is a chance for an increase in the number of samples with unmonitored lines that violate the line limit rating.

However, there is benefit of choosing higher loading threshold for branch classification because it will reduce the number of critical lines in the subset of monitored lines. Looking at Fig. 15, we can clearly see there is a downward correlation trend between loading threshold and the amount of time it took to solve 2000 ROPF samples. ROPF performs better at every threshold comparing to the baseline time of full OPF formulation problems.

TABLE II
Total time to solve 2000 ROPF problems at multiple loading thresholds using GNN prediction. Represent as a percentage of normal OPF problem.

| Model    |     | Time (s) | Time (%) |
|----------|-----|----------|----------|
| Full OPF |     | 145.87   | 100.00   |
| ROPF     | 70% | 130.88   | 89.72    |
|          | 75% | 126.57   | 86.77    |
|          | 80% | 123.39   | 84.59    |
|          | 85% | 122.38   | 83.90    |
|          | 90% | 121.05   | 82.99    |
|          | 95% | 118.97   | 81.56    |

From TABLE II, at the highest threshold of 95%, it only takes 118.97 seconds to solve 2000 ROPF samples compare to the baseline time of 145.87 seconds for full OPF, a saving of almost 18% in computing time. However, when the 95% threshold was chosen in TABLE III, it has the highest prediction error with only 12% of the total number of lines are monitored. We saw that almost 41% of the ROPF samples contain at least one branch that violates the line rating limit. These lines belong in the subset of unmonitored lines. We concluded previously that a high number of type 2 prediction errors is associated with a high number of ROPF samples with unmonitored lines that violate the line rating limit. Then, it is not recommended to set a 95% loading threshold for branch classification. Based on our observation, setting the threshold at 85% of line capacity produces the best result in terms of time saving and accuracy. At 85%, the computation time saving is 16%, a 2% reduction compared to 18% at 95%, but only 0.4% of the ROPF samples will contain at least one branch that violates the line rating limit. This is a significant gain in terms of accuracy compared to minimal loss in time saving.

TABLE III
Comparison of average measurement metrics of 2000 ROPF testing samples at multiple loading threshold using GNN model for branch classification.

| Threshold | Time (%) | % of Samples Over Limit | % of Lines Monitored | Prediction Error (%) |
|-----------|----------|-------------------------|----------------------|----------------------|
| 70%       | 89.72    | 0.2                     | 27.27                | 1.47                 |
| 75%       | 86.77    | 0.2                     | 23.66                | 2.23                 |
| 80%       | 84.59    | 0.25                    | 20.55                | 1.85                 |
| 85%       | 83.90    | 0.4                     | 17.67                | 2.26                 |
| 90%       | 82.99    | 9.65                    | 15.17                | 2.33                 |
| 95%       | 81.56    | 41.35                   | 11.99                | 2.71                 |

An interesting observation can be made regarding the figure below. At 85% loading threshold, only 0.4% or eight samples contain solution with at least one branch that violate line limit restriction. Fig. 16 depicts those eight samples. For these samples, a small reduction, less than 1%, in total generation cost is achieved at the cost of violating line limit restrictions.

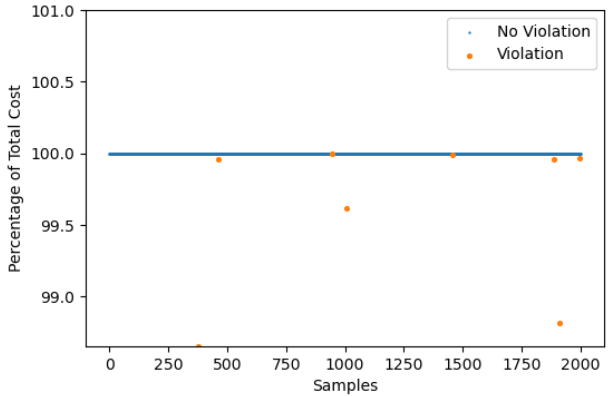

Fig. 16. Total generation cost of ROPF solutions at 85% loading threshold.

## V. Conclusion

From the error prediction result of all three ML models on the testing data set, we can conclude that GNN is the best model for classifying branches that will likely be overloaded or congested given a certain load profile. Having the lowest error prediction rate across all thresholds leads to additional saving in computing time of the proposed ROPF model. Choosing the right loading threshold for branch classification is another important step since accuracy matters as much as speed. At 85% threshold, predictions from the GNN model produce only 0.4% of solved ROPF cases that have at least one branch violating rating limit with a 16% reduction in computing time. This is a considerable improvement in terms of resources saving or computational cost.

In our research, we believe that model selection is an important step for the training of ML algorithm. GNN model, with its preservation of node features, edge features, and topology of the network during the training process, is a superior choice comparing to NN and CNN. Building and training GNN models can be difficult, requiring large amount of data, and computationally expensive, but the reward is significant. Prediction can be made quickly once the GNN model has been trained. Applications of GNN in power systems are still being explored and researched. Regardless, the potentials of using GNN model to build and improve upon previous power systems technologies are substantial.

## VI. Future Work

The proposed GNN model shows significant benefit in computing time saving for calculating OPF for real-time economic dispatch. With our experience in building and training GNN model, we believe that we can produce better results with higher computing time saving using a more complex and practical OPF model on much larger power systems. Due to the adaptability nature of GNN model, accurate predictions can still be made with different topologies of the power system network. We plan to develop a GNN model that quickly screens *N*-1 contingencies and detects lines that will violate line limit rating in the post-contingency situation.